\documentclass[conference]{IEEEtran}
\IEEEoverridecommandlockouts
\usepackage{algorithm}
\usepackage{algorithmic}
\usepackage[mode=buildnew]{standalone}
\usepackage{amsmath,mathtools,nccmath}
\usepackage{siunitx}
\usepackage{cite}
\usepackage{graphicx}
\usepackage{dblfloatfix}
\usepackage{booktabs}
\usepackage{multirow}
\usepackage{enumitem}
\usepackage{mathtools}
\usepackage{amssymb}
\usepackage{gensymb}
\usepackage{float}
\usepackage{array}
\usepackage{textcomp}
\usepackage{xcolor}
\usepackage{booktabs}
\usepackage{multirow}
\usepackage{enumitem}
\usepackage{mathtools}
\usepackage{url}
\usepackage{soul}
\usepackage{fancyhdr}
\def\BibTeX{{\rm B\kern-.05em{\sc i\kern-.025em b}\kern-.08em
    T\kern-.1667em\lower.7ex\hbox{E}\kern-.125emX}}
    
\begin{document}

\title{Real-Time Simulation of a Resilient Control Center for Inverter-Based Microgrids\\

\thanks{
© 20XX IEEE.  Personal use of this material is permitted.  Permission from IEEE must be obtained for all other uses, in any current or future media, including reprinting/republishing this material for advertising or promotional purposes, creating new collective works, for resale or redistribution to servers or lists, or reuse of any copyrighted component of this work in other works.

This work is supported in part by the National Science Foundation (NSF) under award ECCS-1953213, in part by the State of Virginia’s Commonwealth Cyber Initiative (www.cyberinitiative.org), in part by the U.S. Department of Energy's Office of Energy Efficiency and Renewable Energy (EERE) under the Solar Energy Technologies Office Award Number 38637 (UNIFI Consortium led by NREL), and in part by Manitoba Hydro International. The views expressed herein do not necessarily represent the views of the U.S. Department of Energy or the United States Government.
}
}

\author{
\IEEEauthorblockN{Milad Beikbabaei, \emph{Graduate Student Member, IEEE}, and Ali Mehrizi-Sani, \emph{Senior Member, IEEE}}
\IEEEauthorblockA{The Bradley Department of Electrical and Computer Engineering\\
Virginia Polytechnic Institute and State University, Blacksburg, VA 24061 \\
e-mails: \{miladb, mehrizi\}@vt.edu}
}

\maketitle
\begin{abstract}

The number of installed remote terminal units (RTU) is on the rise, increasing the observability and control of the power system. RTUs enable sending data to and receiving data from a control center in the power system. 
A distribution grid control center runs distribution management system (DMS) algorithms, where the DMS takes control actions during transients and outages, such as tripping a circuit breaker and disconnecting a controllable load to increase the resiliency of the grid. 
Relying on communication-based devices makes the control center vulnerable to cyberattacks, and attackers can send falsified data to the control center to cause disturbances or power outages. Previous work has conducted research on developing ways to detect a cyberattack and ways to mitigate the adverse effects of the attack. This work studies false data injection (FDI) attacks on the DMS algorithm of a fully inverter-based microgrid in real time. 
The fully inverter-based microgrid is simulated using an RTDS, an amplifier, an electronic load, a server, a network switch, and a router. 
The DMS is integrated into the server codes and exchanges data with RTDS through TCP/IP protocols. Moreover, a recurrent neural network (RNN) algorithm is used to detect and mitigate the cyberattack. The effectiveness of the detection and mitigation algorithm is tested under various scenarios using the real-time testbed.

\end{abstract}


\section{Introduction}

Communication devices such as remote terminal units (RTU) are widely used in the power system, helping to improve the observability and controllability of the power system. 
An RTU can exchange data with a control center, where it transmits electrical measurements, such as voltage and current, to the control center and receives control commands, such as circuit breaker trips and power generation set points of distributed energy generations (DER) from the control center~\cite{review_CPS_smartgrid}. 
In a distribution grid, the control center runs steady-state algorithms such as load flow and state estimation. 
Moreover, it runs distribution management system (DMS) algorithms to maintain the grid voltage and frequency within a nominal range using the received measurements from RTUs. 

Using communications can open rooms for intruders to launch a cyberattack by falsifying data packets being sent to the control center. If the falsified data are not detected, it can prevent load flow and state estimation from converging or even result in unnecessary control action, causing overvoltage and overcurrent~\cite{review_CPS_smartgrid,ML_based_IDS_review_SG}.
Numerous cyberattacks have been reported during the past few years on the power system~\cite{review_case_cyber_security}. One of the largest successful cyberattacks occurred in 2015, causing a major blackout in Ukraine and leaving about 225,000 thousand Ukrainians without electricity~\cite{parvania_realtime}. 
Moreover, more communications have been installed in the power system, increasing its vulnerability against cyberattacks.
As a result, extensive research has been conducted to simulate cyberattacks and study the impact of a successful attack on the power system.

Previous work has developed various platforms to study the cybersecurity of the power system.
One way to study cybersecurity is using a cosimulation platform to simulate the power system and communication software together, where NS2, NS3, OMENT++, MATLAB Simulink, and OPNET software are used for the communication system, and PSCAD, ETMP/RV, ETMP/ATP, PSS/E, and PSLF software are used for the power system~\cite{Co_simulation_GLABD_NS3, Milad_Ashwin_cosim}.
Another solution is using a real-time testbed to study cybersecurity~\cite{parvania_realtime}. 
In a real-time simulation, every 1~s of simulation takes exactly 1~s in real life. 
As a result, physical hardware can be connected to a real-time simulator (RTS) and simulate part of the simulation, offering a cost-effective way to study the power system more accurately. 
Reference~\cite{parvania_realtime} develops a real-time testbed for studying the cyber vulnerability of the distribution grid using Opal-RT.
A real-time testbed is developed using Opal-RT and network switches for training SCADA operators of a distribution system~\cite{realtime_operating_applied_science}.
Reference~\cite{milad_5G_testbed} develops a real-time testbed using 5G communications for studying the integration inverter-based of the future 5G-enabled grid.

Previous work proposes methods to detect the cyberattack and mitigate its negative impact after it has been detected. Machine learning (ML)--based methods show high accuracy for detecting and mitigating various types of cyberattacks such as false data injection (FDI) and denial of service (DoS) attacks~\cite{ML_based_IDS_review_SG, review_DoS_smartgrid_RL}.
Attackers can launch an FDI attack on the measurements being sent to the control center to pass the conventional bad data detection (BDD) algorithm, and~\cite{BDD_CNN_SE} uses a convolutions neural network (CNN) algorithm to detect it.
In~\cite{CNN_recovery_cyber_physical}, CNN and representation-learning are used to detect and mitigate FDI attacks on the power system measurements.
An unsupervised algorithm is developed to detect anomalies in RTU measurements using feature extraction of unlabeled data in~\cite{deep_scalable_unsupervised}.
%
ML algorithms are used in real time applications~\cite{ML_in_realtime,ML_in_realtime_2}, and the effectiveness of neural network methods for FDI attack detection are tested in real time~\cite{parvania_realtime}.
Reference~\cite{milad_FDI_LSTM} uses a long short-term memory (LSTM)--based algorithm to detect and mitigate FDI attacks on the inverter power set points of a fully inverter-based microgrid.
%
A gated recurrent unit (GRU)--based detector is developed to detect electricity theft on DER generation, where a larger real power generation is reported to the utility in~\cite{GRU_RNN_electricity_theft}.
Various deep learning methods are used for detecting electricity theft on metering infrastructure data in~\cite{DL_AMI_GRU}, where GRU shows the best performance.
Previous work has studied ways to detect cyberattacks on the SCADA system and energy management system (EMS); however, very few have studied the detection and mitigation of cyberattacks on DMS algorithms.
Various deep learning methods are used to detect FDI attacks on the SCADA system in~\cite{SCADA_DL}.
Reference~\cite{DMS_blockchain} uses a blockchain-based architecture to enhance cyberattack detection of a DMS system, but it cannot mitigate the attack.

Designing a cyber-resilient DMS is essential for maintaining the voltage and frequency of a fully inverter-based microgrid under FDI attacks. This work proposes a detection and mitigation method for FDI attacks on the DMS algorithm of a fully inverter-based microgrid using a real-time testbed. 
The simulation runs using RTDS. This work has the following features:
\begin{itemize}
\item A real-time testbed is developed to study the cyber vulnerability of fully inverter-based microgrids. 
\item Detection and mitigation are performed using the GRU algorithm in real time.
\item The effectiveness of the detection and mitigation algorithm is tested in real time using RTDS, a four-quadrant amplifier, an electronic load, a network switch, and a server.
\end{itemize}

The real-time testbed setup is discussed in the next section. Section~III discusses the implementation of the microgrid using the testbed. Section~IV discusses cyberattack detection and mitigation, Section~V presents the performance evaluation, and Section~VI concludes the paper.

\section{ Testbed Setup }

\begin{figure}[!t]
\centerline{\includegraphics[width= 0.72\columnwidth]{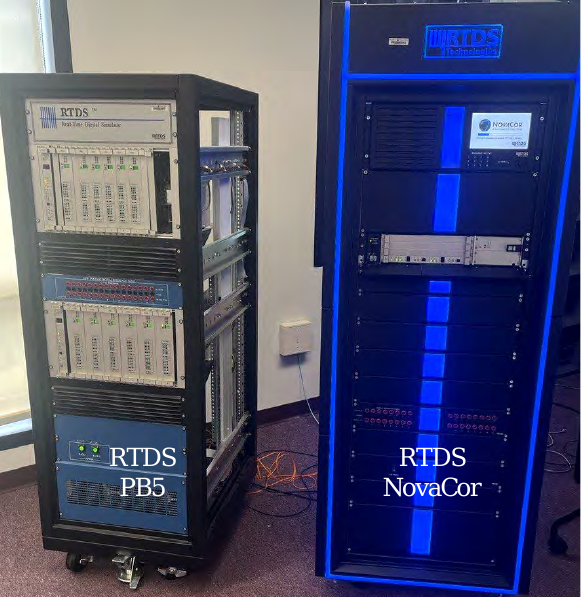}}
\caption{ NovaCor and PB5 racks of the RTDS setup. }
\label{rtdsvt}
\end{figure}

A real-time testbed has been set up at Virginia Tech (VT), consisting of three racks of RTDS, where it has two racks of PB5 technology and one rack of NovaCor technology shown in Fig.~\ref{rtdsvt}.
Each rack of RTDS can simulate a limited number of nodes, and one way to simulate a larger grid is to use all three racks together using a global hub. The global hub ensures the synchronization of all three racks and connects all racks together using optic fiber. Both PB5 and NovaCor have digital and analog inputs and outputs that facilitate the exchange of voltage and current with physical hardware. 
Moreover, NovaCor comes with a giga-transceiver network communication card (GTNET). GTNET is used for implementing network communication, such as transmission control protocol (TCP), user datagram protocol (UDP), MODBUS, and distributed network protocol 3 (DNP3). Furthermore, GTNET supports ethernet cables,  optic fibers, and serial cables.

\setlength{\textfloatsep}{5pt}

\begin{figure}[!t]
\centerline{\includegraphics[width= 0.72\columnwidth]{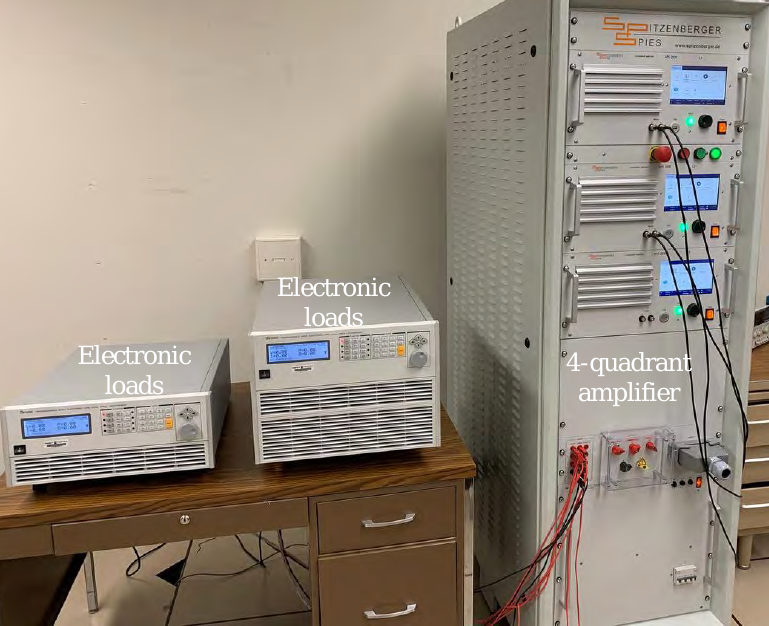}}
\caption{ A four-quadrant amplifier and electronic loads. }
\label{amplifier_load}
\end{figure}

Two programmable AC/DC loads are connected to a three-phase four-quadrant amplifier, as shown in Fig.~\ref{amplifier_load}.
The maximum real power of the smaller electronic load is 1.8~kW, and the larger one has 4.5~kW maximum real power.
The real and reactive power of the electronic loads can be adjusted using serial communication. 
The amplifier can inject up to 2.5~kVAr per phase, and each phase is controlled individually. The amplifier can inject and consume both real and reactive power, its rise time is under 5~us, and have an optical fiber supporting Aurora protocol that can adjust its output voltage in real time.

\setlength{\textfloatsep}{5pt}

\section{ Implemented Microgrid }

This section discusses the implementation of the microgrid and how data is exchanged in the testbed.

\subsection{RTDS Implementation}
\setlength{\textfloatsep}{5pt}
\begin{figure}[!t]
\centerline{\includegraphics[width= 0.95\columnwidth]{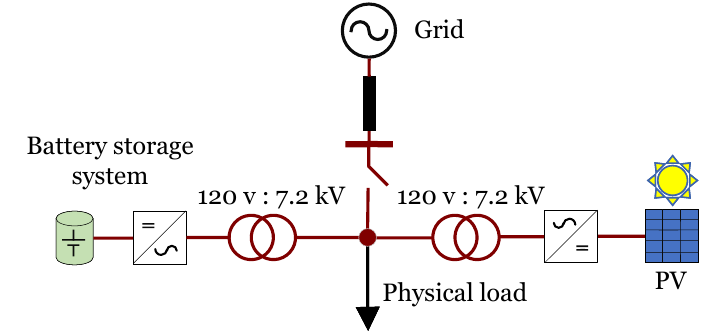}}
\caption{ Study microgrid. }
\label{icdreg_powersystem}
\end{figure}

Fig.~\ref{icdreg_powersystem} shows the microgrid diagram implemented in this work. 
The microgrid consists of a photovoltaic (PV) unit, a battery energy storage system (BESS) unit, a physical load, and a voltage source.
The microgrid is modified using RTDS \verb!RaspberryIsland! example~\cite{RTDS_Example}, by replacing the load in the RTDS with the four-quadrant amplifier model. 
The four-quadrant amplifier and the electronic load represent the physical load. 
The load is modeled using a current source, where the amplifier gets voltage from the RTDS and sends the electronic load current to the GTNET using an optic fiber to minimize the communication delay. 
RTDS scales the received current by a factor of 100.
The electronic load simulates both critical and controllable loads, where the critical load is 600~W, and the controllable load is 800~W.
RTDS sends real power and voltage measurements during the simulation using a GTNET card and TCP/IP protocol to a PC, which runs the server Python code.
A DMS algorithm is integrated within the server code, where it can adjust the real power of the electronic load using serial communications.

The PV unit is modeled in the RTDS, where it is connected to an inverter that can produce up to 250~kW when the insolation of the sun is 1000 and operating in the grid-following mode. Changing the insolation changes the PV generation~\cite{PV_shading}.
The PV panel has 36 cells, where the cell open circuit voltage is 21.7~V, the cell short circuit current is 17.4~A, the voltage at the maximum power is 17.4 V, and the current at maximum power is 3.05~A.
The PV inverter is connected to a transformer through an RL filter, where the filter resistance is 10~$\micro \Omega$ and its reactance is 30~$\micro$H. The primary and secondary voltages of the transformer are 120~V and 7.2~kV. The series resistance and inductance of the transformer are 30~$\micro \Omega$ and 137.5~mH.

The BESS unit is modeled in the RTDS, where it is connected to an inverter that can produce up to 100~kW. 
It has a lithium-ion battery, which has 150 series and 500 parallel slacks. The battery type is set to Min/Rincon-Mora, and the initial state of charge (SoC) of the battery is set to 35\%.
The BESS inverter is connected to a transformer through an RL filter, where the filter resistance is 10~$\micro \Omega$ and its reactance is 30~$\micro$H. The primary and secondary voltage of the transformer is 120~V and 7.2~kV. The series resistance and inductance of the transformer are 30~$\micro \Omega$ and 137.5~mH.
An ideal voltage source represents the grid in the RTDS.

\subsection{ Data Exchange Path}

\setlength{\textfloatsep}{5pt}
\begin{figure}[!t]
\centerline{\includegraphics[width= 0.92\columnwidth]{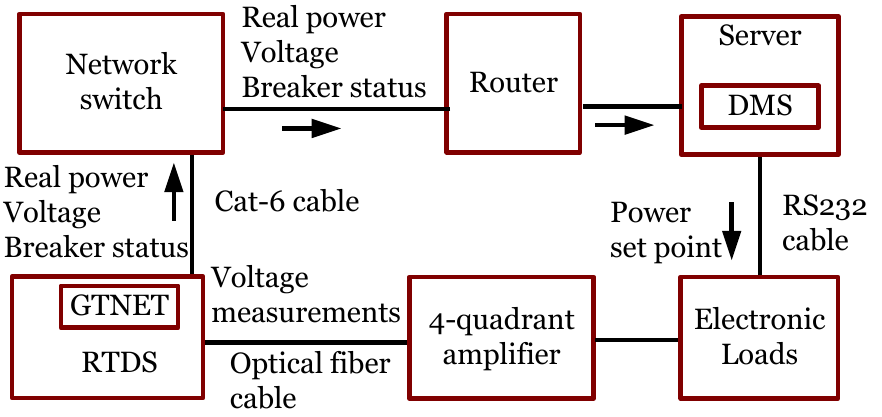}}
\caption{ Data exchange path of the testbed. }
\label{VT_TEST_BED_Cyber_layer_ver2}
\end{figure}

Fig.~\ref{VT_TEST_BED_Cyber_layer_ver2} shows how data is exchanged in the testbed. 
RTDS simulates the microgrid, sending the voltage measurements to the amplifier and receiving current output measurements of the amplifier using the optic fiber in real time.
RTDS sends the real power measurements, voltage measurements, and grid breaker status in real time to a control center using the GTNET card and TCP/IP protocol to a network switch.
The network switch and router send received data to the DMS server.
The PV and BESS real power measurements are scaled down by a factor of 100 to match the load measurements before sending them to the server.
The DMS algorithm disconnects the controllable load in the islanding mode of operation if the SoC of the battery is less than 50\% to have enough power to provide the critical for a longer period of time. As soon as the SoC becomes greater than 50\% or the grid gets reconnected, the DMS reconnects the controllable load.
DMS adjusts the real power of the electronic load using serial communication in real time.

\section{Cyberattack Detection and Mitigation}

This section introduces GRU and then describes the cyberattack detection and mitigation method.

\subsection{Gated Recurrent Unit (GRU) Basics}

\begin{figure}[!t]
\centerline{\includegraphics[width= 0.8\columnwidth]{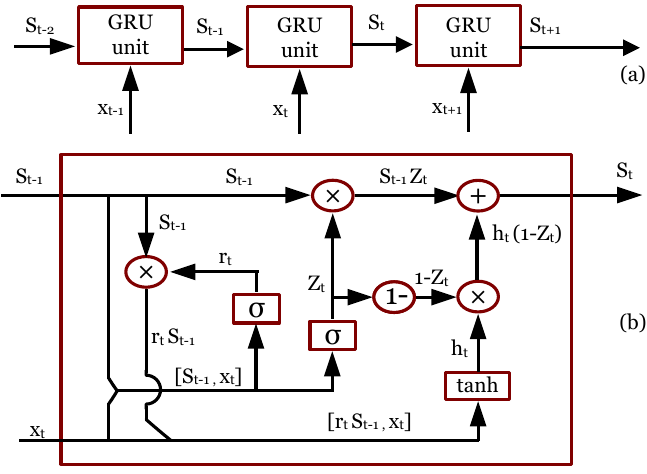}}
\caption{ A GRU unit: (a) in the time domain (b) gates structure of a GRU unit. }
\label{GRU_unit}
\end{figure}

A recurrent neural network (RNN) algorithm is a type of neural network that is widely used in the prediction of time series data, such as weather measurements, where having a memory of the previous data helps with the accurate prediction of the next time step. 
A gated recurrent unit (GRU) is a type of RNN. Basic RNN cannot deal with long-term dependencies of past data; however, GRU can deal with long-term dependencies using reset and update gates. 
GRU is faster than LSTM due to the fewer number of gates. Fig.~\ref{GRU_unit}(b) shows the GRU unit, where it uses reset and update gates. Inputs of the GRU unit are the previous time step output $s_\text{t-1}$ and the new input data $x_\text{t}$, and the output of GRU is $s_\text{t}$~\cite{GRU_basics}.

\setlength{\textfloatsep}{5pt}

The reset gate decides how much of the previous time step output needs to be kept, and the update gate decides how much of the calculated hidden state needs to be forgotten. Moreover, activation functions are used to map numbers into a certain range. The $\sigma$ activation function maps the data to a number between $0$ and $1$, and the $\tanh$ activation function maps the data to a number between $-1$ and $1$. 
The reset gate output $r_\text{t}$ is

\begin{align}
\begin{split}
    r_\text{t} = \sigma { (\: W_\text{r} \: [x_\text{t}, s_\text{t-1}] + b_\text{r}) } ,
    \end{split}
    \label{eq1_GRU}
\end{align}
where $r_\text{t}$ is calculated by passing reset weights vector $W_\text{r}$,  reset bias vector $b_\text{r}$, $x_\text{t}$, and $s_\text{t-1}$ to a $\sigma$ activation function. The hidden state $h_\text{t}$ is

\begin{align}
\begin{split}
    h_\text{t} = \tanh { (\: W_\text{h} \: [x_\text{t}, s_\text{t-1}] + b_\text{h}) } ,
    \end{split}
    \label{eq2_GRU}
\end{align}
where $h_\text{t}$ calculates how much of the new information can be passed to the update gate. It is calculated by passing the weights vector $W_\text{h}$, bias vector $b_\text{h}$, $x_\text{t}$, and $s_\text{t-1}$ through a $\tanh$ activation function. The update gate value $z_\text{t}$ is

\begin{align}
\begin{split}
    z_\text{t} = \sigma { (\: W_\text{z} \: [x_\text{t}, s_\text{t-1}] + b_\text{z}) } ,
    \end{split}
    \label{eq3_GRU}
\end{align}
where $z_\text{t}$ is calculated by passing the weights vector $W_\text{z}$, bias vector $b_\text{z}$, $x_\text{t}$, and $s_\text{t-1}$ through a $\sigma$ activation function. The output state $s_\text{t}$ is

\begin{align}
\begin{split}
    s_\text{t} = (1-z_\text{t})\: h_\text{t} + z_\text{t}\:s_\text{t-1} ,
    \end{split}
    \label{eq4_GRU}
\end{align}
where $s_\text{t}$ is calculated using $z_\text{t}$, $h_\text{t}$, $z_\text{t}$, and $s_\text{t-1}$.
Fig.~\ref{GRU_unit}(a) shows how GRU outputs are used in the next time step for three consecutive time steps.

\subsection{ Proposed Method}

\setlength{\textfloatsep}{5pt}
\begin{figure}[!t]
\centerline{\includegraphics[width= 0.95\columnwidth]{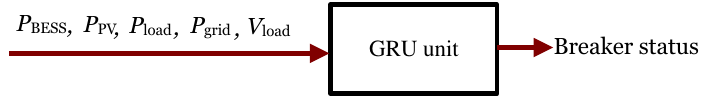}}
\caption{ Inputs and outputs of the GRU block. }
\label{GRU_detection}
\end{figure}

The DMS has a GRU block to detect and mitigate FDI attacks on the grid breaker status using the real power of the BESS, the real power of the PV unit, the real power of the grid, the real power of the load, and the voltage of the loads.
The GRU block is shown in Fig.~\ref{GRU_detection}.
The GRU block output is a binary value, where 0 means it is in the grid mode and 1 means it is in the islanded mode.
The estimated grid breaker status is the GRU block outputs, as shown in Fig.~\ref{GRU_detection}. 
Each GRU unit uses 5 inputs, which helps keep the accuracy high even if one of the measurements is noisy due to the communication noise or sensor malfunction.
If the GRU block estimation and the received breaker status are different, the attack is detected. The GRU block estimated value is replaced with the falsified status to mitigate the cyberattack. A successful cyberattack can cause outages for the dispatchable load in the grid-connected mode or drain the battery sooner by not disconnecting the controllable load in the islanded mode. The accuracy of the trained GRU is 94.91\%.

\subsection{ Training and Parameters}

A dataset is created using both the islanded and grid-connected modes data.
In the grid-connected mode, a dataset is generated by changing the PV generation from 0 to 2400~W with steps of 200~W and changing the BESS real power set point from 100~W to 700~W with steps of 100~W, resulting in 91 cases.
In the islanded mode, a dataset is generated by changing the PV generation from 0~W to 2500~W with steps of 30~W, resulting in 84 cases.
70\% of the dataset is used for training, and 30\% of the dataset is used for testing.
The GRU is implemented using the Keras library and Python 3.10~\cite{GRU_keras_tf}.
The GRU uses five measurements as inputs to predict the output, which is the breaker status, as shown in Fig.~\ref{GRU_detection}. Furthermore, GRU has an input and an output layer.
%
Table~\ref{GRU_parameters} shows the implemented GRU parameters, such as activation function, number of neurons, layer types, learning rates, loss functions, and the selected optimizer.
%
GRU algorithm is trained for 200 epochs, the initial bias value is 0. 
The number of training epochs and optimizer learning rates are selected through trial and error.

\begin{table}[!t]
    \footnotesize\centering
    \caption{GRU Parameters }
    \label{GRU_parameters}
    \setlength{\tabcolsep}{0.8mm}\begin{tabular}{llllll}
        \toprule
        \multicolumn{2}{c}{\textbf{GRU Parameters}} & \multicolumn{2}{c}{\textbf{Input Layer}} & 
        \multicolumn{2}{c}{\textbf{Output Layer}} 
        \\
        \cmidrule(r){1-2}\cmidrule(r){3-4}\cmidrule(r){5-6}
    \textbf{ Parameter } & \textbf{ Value } & \textbf{ Parameter } & \textbf{ Value } & \textbf{ Parameter } & \textbf{ Value } \\ 
        \cmidrule(r){1-6}
        \multirow{2}{*}{\shortstack{Learning \\rate}} & \multirow{2}{*}{0.001} &  \multirow{2}{*}{\shortstack{Number of\\ neurons}} & \multirow{2}{*}{50} & \multirow{2}{*}{\shortstack{Number of\\ neurons}} & \multirow{2}{*}{1}\\ \\
        \cmidrule(r){1-6}
        \multirow{2}{*}{\shortstack{Loss \\ function}} & \multirow{2}{*}{MSE} &  \multirow{2}{*}{\shortstack{Layer type}} & \multirow{2}{*}{GRU} & \multirow{2}{*}{Layer type} & \multirow{2}{*}{\shortstack{Dense }}\\ \\
        \cmidrule(r){1-6}
        \multirow{2}{*}{Optimizer} & \multirow{2}{*}{Adam} &  \multirow{2}{*}{\shortstack{Activation \\ function}} & \multirow{2}{*}{$\tanh$} & \multirow{2}{*}{\shortstack{Activation \\ function}} & \multirow{2}{*}{sigmoid}\\ \\
        \bottomrule
    \end{tabular}
\end{table}

\section{Performance Evaluation}

This section discusses the simulation results for FDI cyberattacks both in the grid-connected and islanded modes of operation.

\setlength{\textfloatsep}{5pt}
\begin{figure}[!t]
\centerline{\includegraphics[width= 0.92\columnwidth]{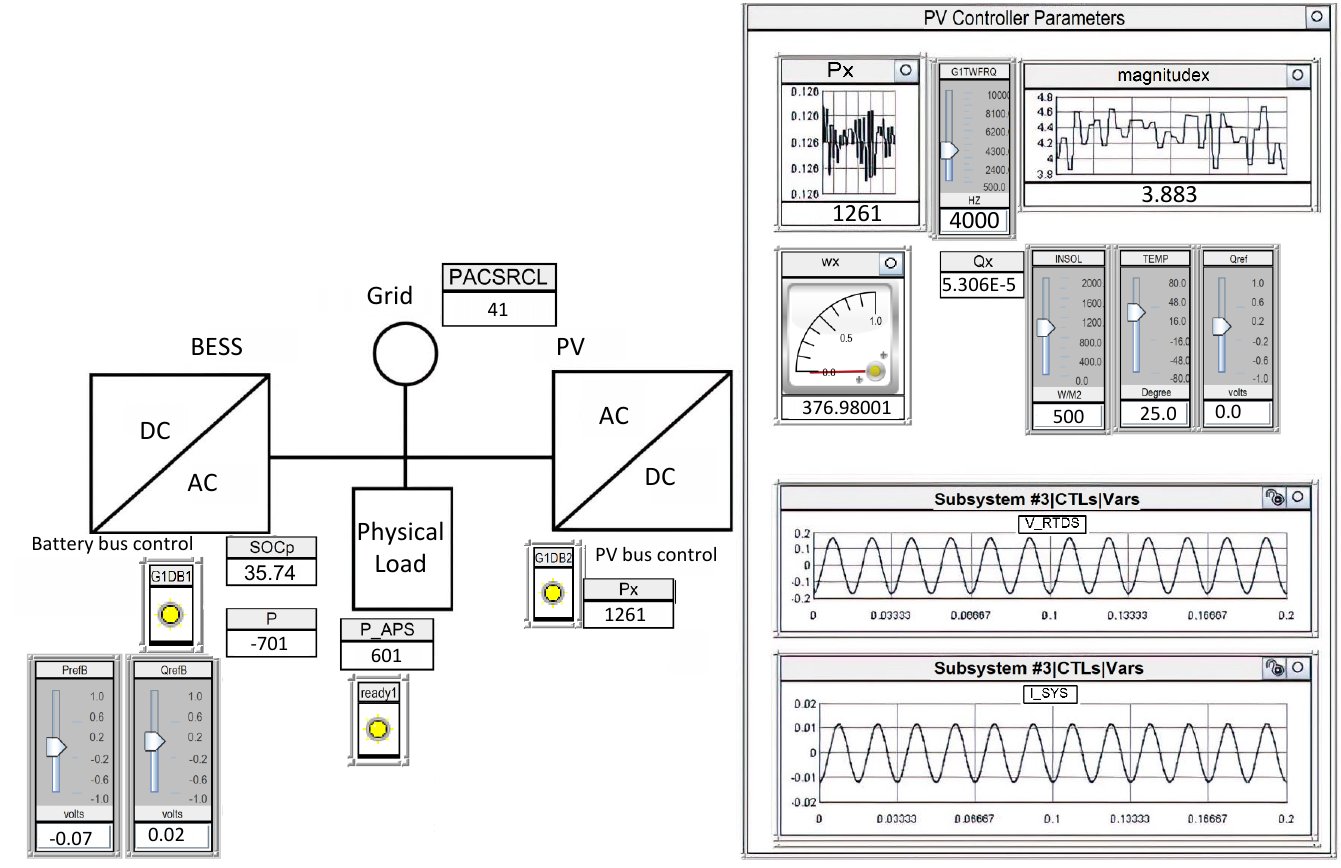}}
\caption{ RTDS runtime for a successful attack in the grid-connected mode. } 
\label{Islanded_Insol_500_horizental}
\end{figure}

\setlength{\textfloatsep}{5pt}
\begin{figure}[!t]
\centerline{\includegraphics[width= 0.82\columnwidth]{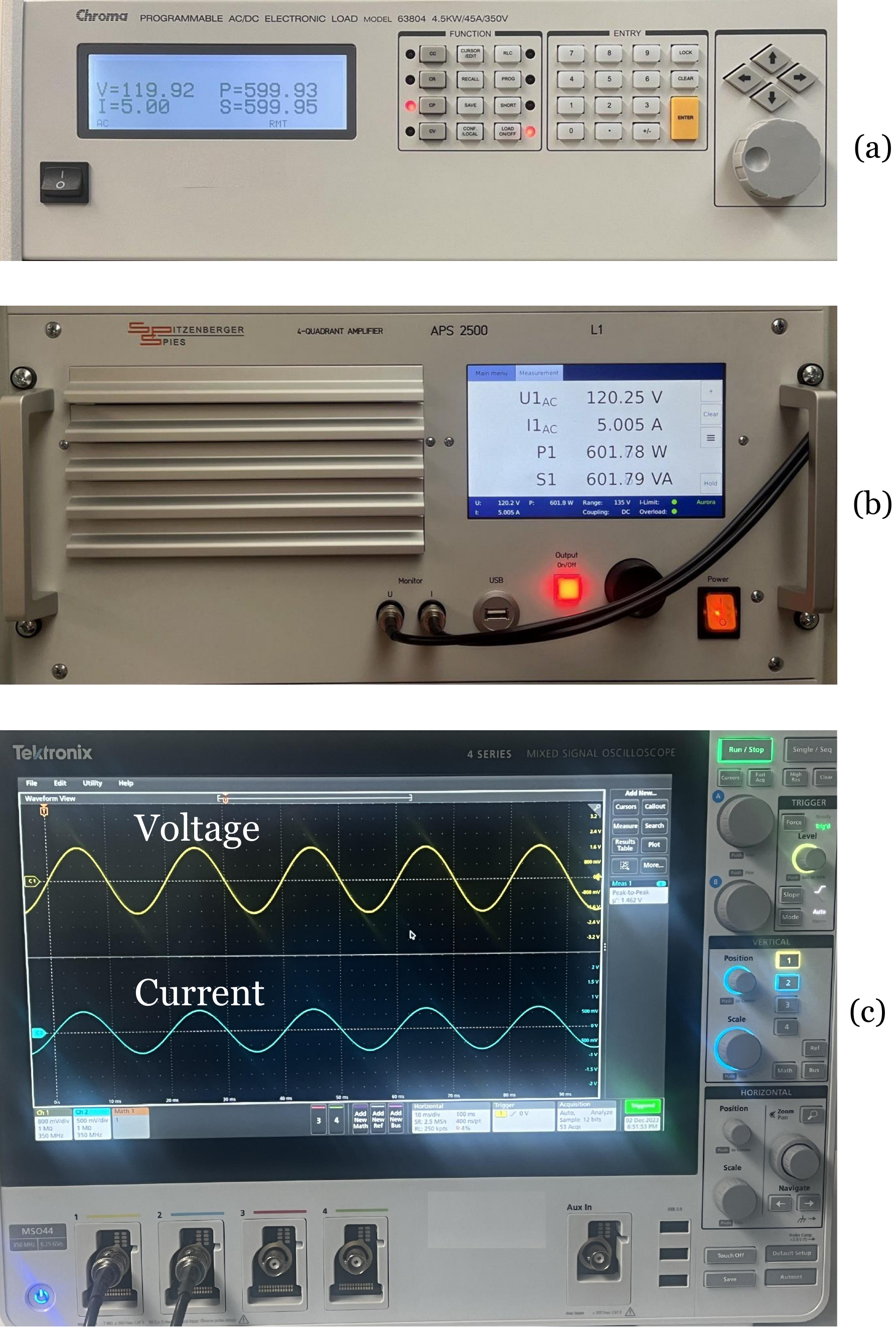}}
\caption{ A successful cyberattack in the grid-connected mode of microgrid: (a) electronic load panel, (b) amplifier panel, and (c) voltage and current of the amplifier. }
\label{islanded_all_merged}
\end{figure}

\subsection{FDI Attack in the Grid-Connected Mode}

In this scenario, the grid circuit breaker is closed, but attackers falsify the circuit breaker status data being sent to the control center. Right before launching the FDI attack, the battery SoC is 35.74\%, and the light insolation is set to 500. 
Without detection and mitigation algorithms, the DMS disconnects the controllable load since the BESS SoC is less than 50\%; however, the FDI attack is detected using the proposed method, and the controllable load does not experience outages.

Fig.~\ref{Islanded_Insol_500_horizental} shows the RTDS runtime of a successful attack, without using the proposed detection and mitigation method, where the controllable load is disconnected, and the load is consuming 600~W. The real power of the amplifier, the BESS, the PV unit, and the grid are shown in Fig.~\ref{Islanded_Insol_500_horizental}.
Fig.~\ref{islanded_all_merged}(a) shows the voltage, current, real power, and apparent power of the electronic load in a successful FDI attack where it is consuming 599.93~W.
Fig.~\ref{islanded_all_merged}(b) shows the voltage, current, real power, and apparent power of the amplifier where it is injecting 601.79~W to the load. The cable connecting the electronic loads to the amplifier has a 1.86~W power loss.
Fig.~\ref{islanded_all_merged}(c) shows the voltage and current of the amplifier when the controllable load is disconnected.

\subsection{FDI Attack in the Islanded Mode}

\setlength{\textfloatsep}{5pt}
\begin{figure}[!t]
\centerline{\includegraphics[width= 0.92\columnwidth]{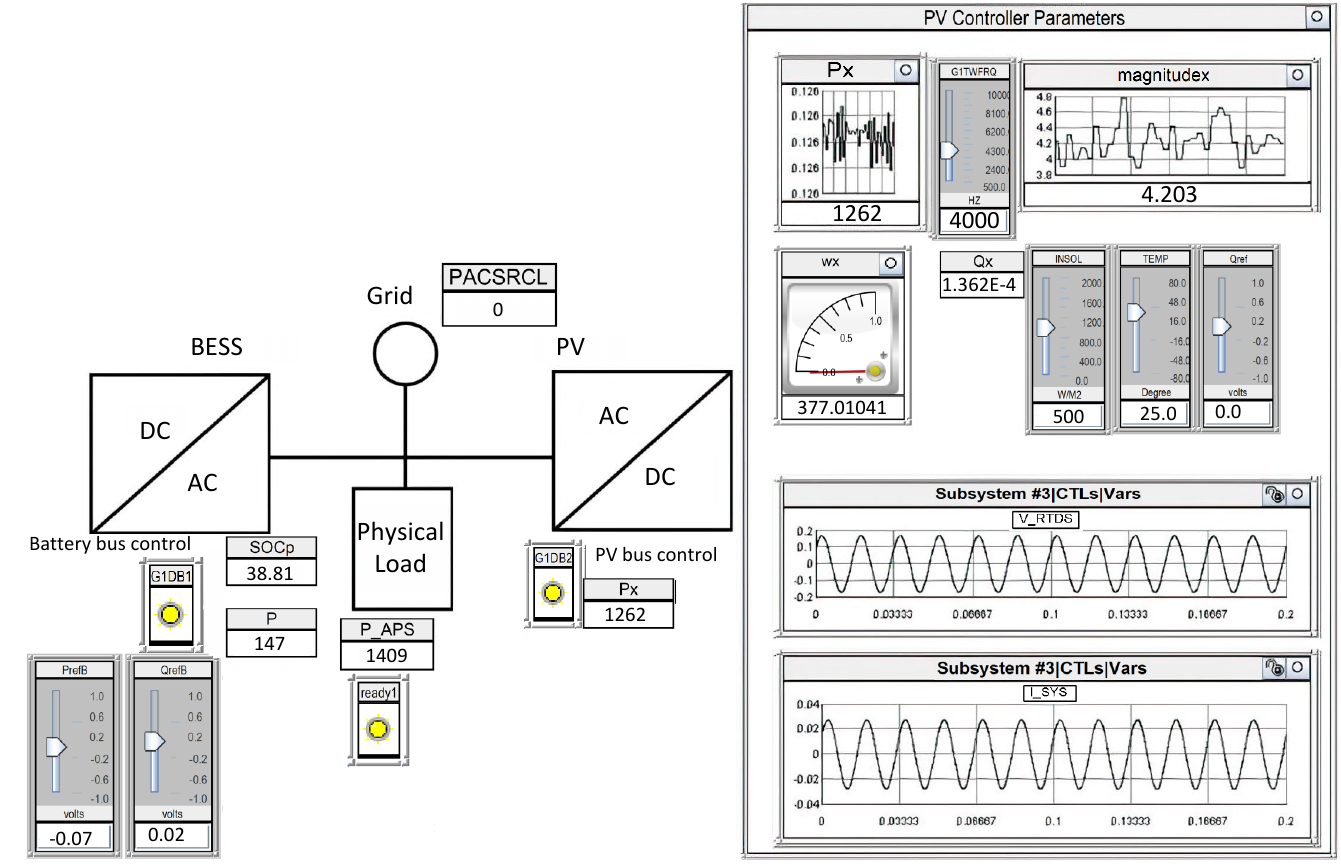}}
\caption{ RTDS runtime for a successful attack in the islanded mode.  }
\label{Gridconnected_Insol_500_horizental}
\end{figure}

\setlength{\textfloatsep}{5pt}
\begin{figure}[!t]
\centerline{\includegraphics[width= 0.82\columnwidth]{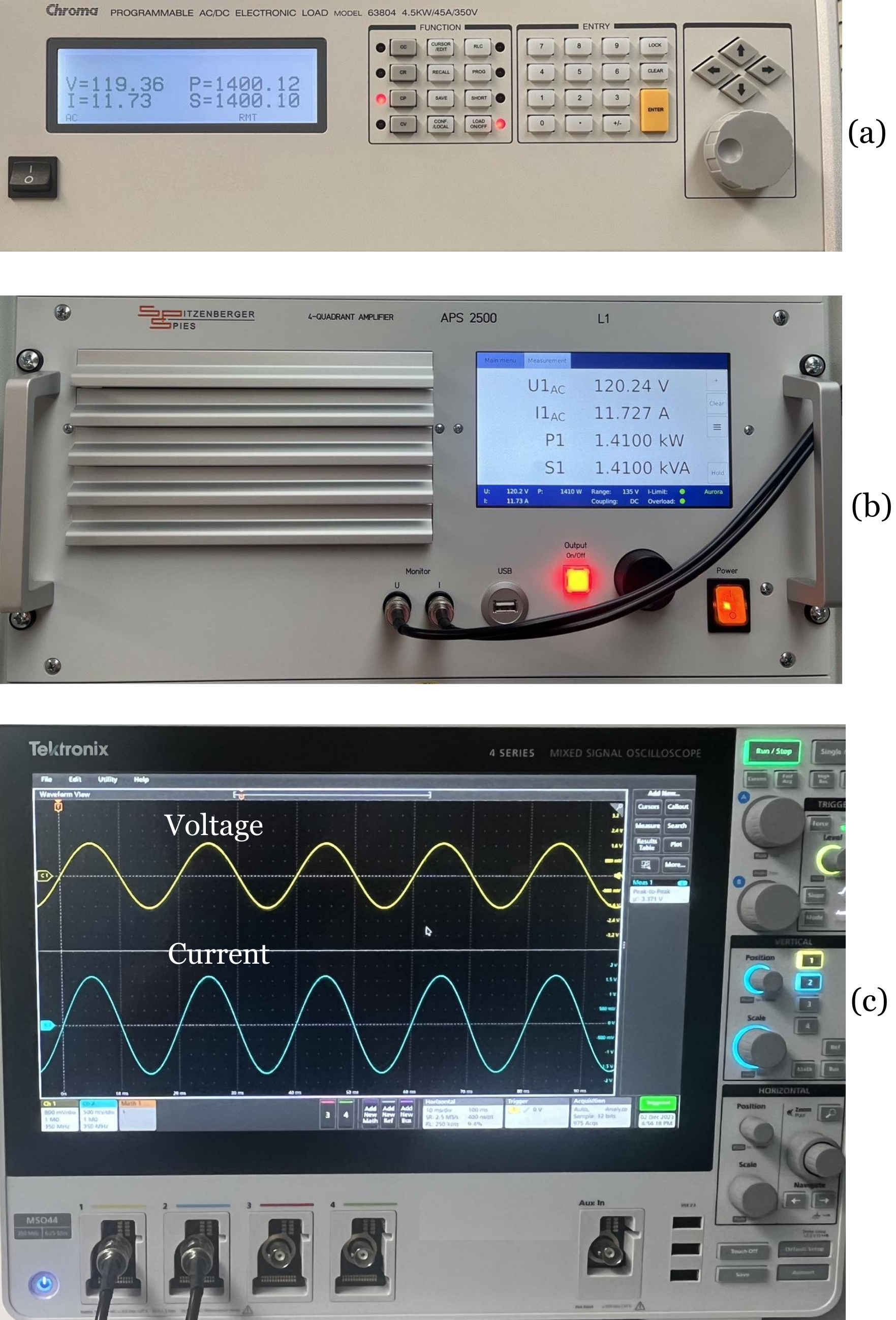}}
\caption{ A successful cyberattack in the islanded mode of microgrid: (a) electronic load panel, (b) amplifier panel, and (c) voltage and current of the amplifier. }
\label{gridconnected_allinone}
\end{figure}

In this scenario, the grid is in the islanded mode, but attackers falsify the circuit breaker status data being sent to the control center. Right before launching the FDI attack, the battery SoC is 38.81, and the light insolation is set to 500. 
Without any detection and mitigation algorithm, the DMS reconnects the controllable load. However, the FDI attack is detected using the proposed method, and the controllable load does not get reconnected, helping the BESS to supply the critical load for a longer period of time.

Fig.~\ref{Gridconnected_Insol_500_horizental} shows the RTDS runtime of a successful attack, without using the detection and mitigation method, where the controllable load is reconnected, and the load is consuming 1400~W. The real power of the amplifier, the BESS, the PV unit, and the grid are shown in Fig.~\ref{Gridconnected_Insol_500_horizental}.
Fig.~\ref{gridconnected_allinone}(a) shows the voltage, current, real power, and apparent power of the electronic load in a successful FDI attack where it is consuming 1400~W.
Fig.~\ref{gridconnected_allinone}(b) shows the voltage, current, real power, and apparent power of the amplifier where it is injecting 1410~W to the load. The cable connecting the electronic loads to the amplifier has a 9.8~W power loss.
Fig.~\ref{gridconnected_allinone}(c) shows the voltage and current of the amplifier when the controllable load is connected.

\section{Conclusion}

This work studies FDI attacks on a DMS system of a fully inverter-based microgrid in real time. The microgrid consists of a PV unit, a BESS unit, and a physical load.
FDI attack is studied both in grid-connected and islanded mode, where a successful attack can disconnect the controllable load in the grid-connected mode and drain the battery sooner by not disconnecting the controllable load in the islanded mode.
The DMS has a GRU block to detect and mitigate the attack. GRU uses the real power of PV, the real power of BESS, the real power of load, the real power of the grid source, and the grid voltage to estimate the grid switch positions.
If the estimated grid breaker position is different from the received position, the attack is detected. Moreover, the attack is mitigated by using the estimated grid breaker position in the DMS algorithm.
The microgrid is tested under FDI attacks for both grid-connected and islanded modes; however, the attack is detected and mitigated using the proposed method.
The implemented real-time testbed in this work facilitates studying FDI attacks on other power system elements such as PV and BESS.

\section*{Acknowledgement}
The authors acknowledge Dr.~A Mohammadhassani for providing the modified RTDS \verb|RaspberryIsland| example.

\bibliography{IEEEabrv,ref}
\bibliographystyle{IEEEtran}

\vspace{12pt}

\end{document}